\documentclass{article}

\usepackage{arxiv}
\usepackage[utf8]{inputenc} 
\usepackage[T1]{fontenc}    
\usepackage{url}            
\usepackage{booktabs}       
\usepackage{amsfonts}       
\usepackage{nicefrac}       
\usepackage{microtype}      
\usepackage{lipsum}		
\usepackage{graphicx}
\usepackage{titlesec}
\usepackage{achemso} 
\setkeys{acs}{usetitle = true}
\usepackage{multirow}
\usepackage{textcomp}

\usepackage{doi}
\usepackage{chngcntr}
\counterwithin{figure}{section}
\usepackage{ragged2e}
\usepackage{anyfontsize}
\usepackage{amsmath}
\usepackage{gensymb}
\usepackage{newtxtext} 
\usepackage{newtxmath} 
\usepackage{multicol}
\usepackage{hyperref}       
\usepackage{caption}        
\usepackage{xcolor}

\usepackage{dcolumn}
\usepackage{bm}
\usepackage{subfigure}
\usepackage{units}
\usepackage{color}
\usepackage[utf8]{inputenc}
\DeclareUnicodeCharacter{2032}{'} 
\DeclareUnicodeCharacter{2212}{-} 

\renewcommand{\footnotesize}{\fontsize{7.68pt}{9.22pt}\selectfont}
\renewcommand{\scriptsize}{\fontsize{6.72pt}{8.06pt}\selectfont}
\renewcommand{\tiny}{\fontsize{6.24pt}{7.49pt}\selectfont}
\renewcommand{\small}{\fontsize{8.64pt}{10.37pt}\selectfont}
\renewcommand{\normalsize}{\fontsize{10.24pt}{12.29pt}\selectfont}
\renewcommand{\large}{\fontsize{13.44pt}{16.19pt}\selectfont}
\renewcommand{\Large}{\fontsize{16.13pt}{19.43pt}\selectfont}
\renewcommand{\LARGE}{\fontsize{19.35pt}{23.22pt}\selectfont}
\renewcommand{\huge}{\fontsize{22.74pt}{27.29pt}\selectfont}
\renewcommand{\Huge}{\fontsize{27.29pt}{32.81pt}\selectfont}

\title{High-Temperature Non-Equilibrium Atom-Diatom Collisional Energy Transfer}

\author{
    \begin{center}
        \textbf{Xiaorui Zhao}\textsuperscript{1,2}, \textbf{Xuefei Xu}\textsuperscript{1,3,*}, and \textbf{Haitao Xu}\textsuperscript{1,2,*}.\\[0.5em]
        \textsuperscript{1} Center for Combustion Energy, Tsinghua University, Beijing 100084, P. R. China \\[0.5em]
        \textsuperscript{2} School of Aerospace Engineering, Tsinghua University, Beijing 100084, P. R. China \\[0.5em]
        \textsuperscript{3} Department of Energy and Power Engineering, Tsinghua University, Beijing 100084, P. R. China \\[0.5em]
        \textsuperscript{*} Co-corresponding author e-mail: xuxuefei@tsinghua.edu.cn, hxu@tsinghua.edu.cn
    \end{center}
}

\titlespacing{\section}{0pt}{\parskip}{-\parskip}
\titlespacing{\subsection}{0pt}{\parskip}{-\parskip}

\begin{document}
\maketitle

\begin{abstract}
The change of the vibrational energy within a molecule after collisions with another molecule plays an essential role in the evolution of molecular internal energy distributions, which is also the limiting process in the relaxation of the gas towards equilibrium. 
Here we investigate the energy transfer between the translational motion and the vibrational motion of the diatom during the atom-diatom collision, the simplest case involving the transfer between inter-molecular and intra-molecular energies. 
We are interested in the situation when the translational temperature of the gas is high, in which case there are significant probabilities for the vibrational energy to change over widely separated energy levels after a collision.
Data from quasi-classical trajectory simulations of the N+N$_2$ system with \textit{ab initio} potential energies suggest that the transition probability dependence on the collisional energy possesses an ``activation-saturation'' behavior and can be described by a simple model. 
The model allows for explicit evaluation of the vibrational state-to-state transition rate coefficients,
from which the evolution of the vibrational energy distribution from any initial conditions can be solved by the master equation approach. An example of the vibrational energy relaxation in the N+N$_2$ system mimicking the gas behind strong shocks in a hypersonic flow is shown and the results are in good agreement with available data. 

\end{abstract}

\renewcommand{\thefigure}{\arabic{figure}}
\justifying

For gases at equilibrium, the internal energy is equally partitioned among the degrees of freedom (d.o.f), i.e., translation, rotation, and vibration of gas molecules, and energies belong to each d.o.f follow the Boltzmann distribution. 
This idealized description well approximates many gas systems and flows encountered. The deviation, however, grows rapidly when the time scale of the condition changing becomes comparable to the thermal relaxation time scale, which is determined by the collisional time among molecules. 
In such non-equilibrium situations, not only the internal energy of the gas system is not equipartitioned among d.o.f's, the energy distributions within some or all d.o.f's are also not Boltzmannian \cite{Park1993,Candler2019}. 
A well-known example is the shocks in hypersonic flows \cite{Park1993,Candler2019,Leyva2017,Singh2018}.
When passing a shock, the kinetic energy in the mean flow converts abruptly to the internal energy in the translational motion (or simply the translational energy) of the gas molecules first, followed by the rotational energy, and eventually to the vibrational energy. 
In this process, the time scales for different d.o.f's to reach equilibrium can differ by several orders of magnitudes, and the maximum differences among temperatures of different internal motions can be as high as $\sim \unit[10^4]{K}$ \cite{Candler2019}.
The key factor in the evolution of this non-equilibrium process is thus the collisional energy transfer among different levels of vibrational energies because of the limiting role of vibrational energy relaxation 
\cite{Park1993,Candler2019,Leyva2017,Singh2018,
Adamovich1995_part1,Adamovich1995_part2,Colonna2008,Colonna2019,
Panesi2013,Kim2013,Panesi2014,
Andrienko2015,Andrienko2016,
Valentini2015,Bender2015,Valentini2016,Macdonald2020,
Campoli2020}.
The seminal Landau-Teller (LT) theory \cite{Landau1936} analyzed the energy transfer
between near-by vibrational energy levels of a diatom during its collision with an atom and obtained the transfer probability, which was later improved quantitatively, e.g., by considering the potential parameters and geometric factors in the Schwartz-Slawsky-Herzfeld (SSH) model \cite{Schwartz1952}.
For gases at low to moderate temperature ($T \sim O(\unit[10^3]{K})$), or when the Mach number of the flow is not too high ($Ma \lesssim 6$), the average kinetic energy change in a collision typically corresponds to the energy difference between neighboring vibrational levels of relevant gas molecules \cite{Adamovich1995_part1}, for which the LT and subsequent theories can provide analytical results with acceptable accuracy.
For hypersonic flows at higher Mach numbers, however, the gas temperature rises after the shock are much higher 
and the probability of energy transfer between two widely-separated vibrational energy levels must be considered, which has been dealt either with simplified inter- and intra- molecular potentials at the cost of accuracy, e.g., in the forced harmonic oscillator (FHO) model \cite{Kerner1958,Adamovich1995_part1,Adamovich1998}, 
or using empirical, many-parameter fits of the energy transfer rates
obtained from molecular dynamics simulations \cite{Billing1976,Billing1992,Esposito2006,Esposito2008,Jaffe2010,Andrienko2015,Hong2020,Hong2021}.

In this letter, we investigate the collisional energy transfer in an atom-diatom bi-molecular system, which is the classical model for this problem \cite{Landau1936, Schwartz1952, Kerner1958, Blais1976, Magill1988, Quemener2007,Zuo2020}.
We observe a universal activation-saturation behavior for the collision-induced energy transfer to higher vibrational levels, which leads to a physics-based semi-analytical model for the rates of vibrational energy transfer.


In the atom-diatom model system, the ``diatom'' is the species whose internal vibrational energy is the quantity of interest 
and the ``atom'' represents the other species of which the internal 
energy distribution is neglected, and the translational energy distributions of the ``diatom'' and the ``atom'' are both Boltzmannian and are at the same temperature. 
It is thus justified to consider the collisions between a ``diatom'' and an ``atom'' from a thermal bath at temperature $T$ 
when studying the collisional energy transfer between different vibrational levels $\nu'$ and $\nu$ of the ``diatoms'', which is characterized by the state-to-state transition rate coefficient $k(\nu' \rightarrow \nu , T)$ ,
with which the evolution of the vibrational energy distribution can be solved through the master equations \cite{Colonna2008,Kim2013,Panesi2013,Andrienko2016}: 
\begin{equation}\label{eq:ME}
    \frac{dn_{\nu}}{dt} = n_{\rm M} \sum_{\nu'}^{}{\lbrack\ k(\nu' \rightarrow \nu , T) n_{\nu'} - k(\nu \rightarrow \nu' , T) n_{\nu}\ \rbrack} 
\end{equation}
where $n_{\nu}$ and $n_{\nu'}$ are the concentration of the ``diatoms'' in the vibrational levels $\nu$ and $\nu'$, 
and $n_{\rm M}$ is the concentration of the ``atoms'', acting as the bath gas. In this approach, the rotational energy of the diatoms does not appear, which is either assumed to be at the Boltzmann distribution with the same temperature as the translational motion and thus is included in the dependence of $k$ on $T$ \cite{Colonna2008}, as also adopted in our treatment later, or is considered to evolve together with the vibrational energy distribution \cite{Kim2013,Panesi2013,Andrienko2016}.

To gain a quantitative understanding of how $k(\nu' \rightarrow \nu , T)$ varies with the temperature $T$ and the energy gap $\nu - \nu'$, 
we simulated the collisions between atom N and molecular N$_2$ using 
the quasi-classical trajectory (QCT) method\cite{Truhlar1979,Bender2015} implemented in the ANT program\cite{Shu2024} 
based on the \textit{ab initio} inter- and intra-molecular potentials \cite{Varga2021},
in which the collision dynamics is treated classically and the initial/final states of the N$_2$ molecule are assigned into discrete vibrational energy states with corresponding quantum numbers $\nu'$/ $\nu$.
In QCT, each trajectory represents the collision dynamics of a N+N$_2$ pair from a specific initial condition, and 
the transition rate coefficients are obtained from statistics over these trajectories (See the Supplementary Material (SM) \cite{SM} for more details). 
The N+N$_2$ system is chosen
because as the major ingredient of air, the energy transfer of nitrogen controls the relaxation of air to equilibrium under abrupt changes like strong shocks. Moreover, compared to oxygen, another important ingredient of air, nitrogen has a higher dissociation energy barrier and a higher electronic excitation
energy level, making it more stable under collisions at high temperatures, hence the collisional energy transfer process is less affected by dissociation and electronic excitation.

\begin{figure}\label{fig:k}
  \centering
  \begin{subfigure}
    \centering
    \includegraphics[width=0.48\columnwidth]{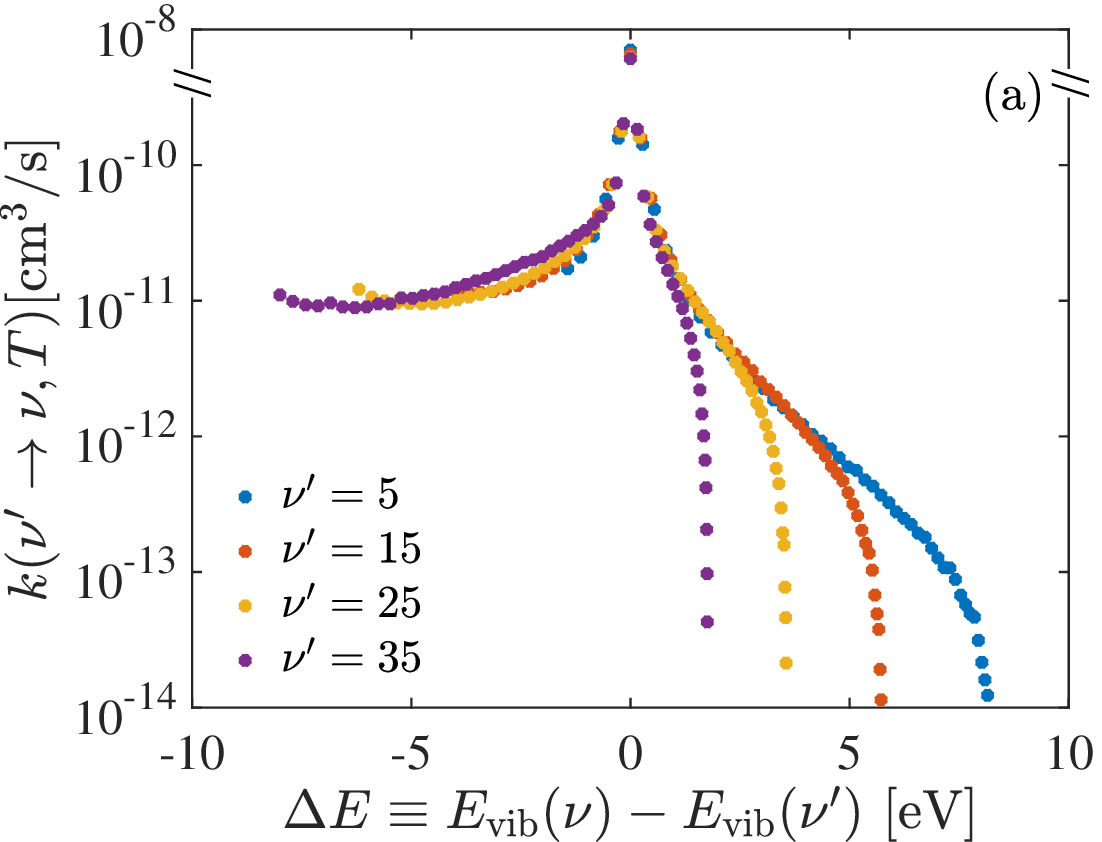}
  \end{subfigure}
  \hfill
  \begin{subfigure}
    \centering
    \includegraphics[width=0.48\columnwidth]{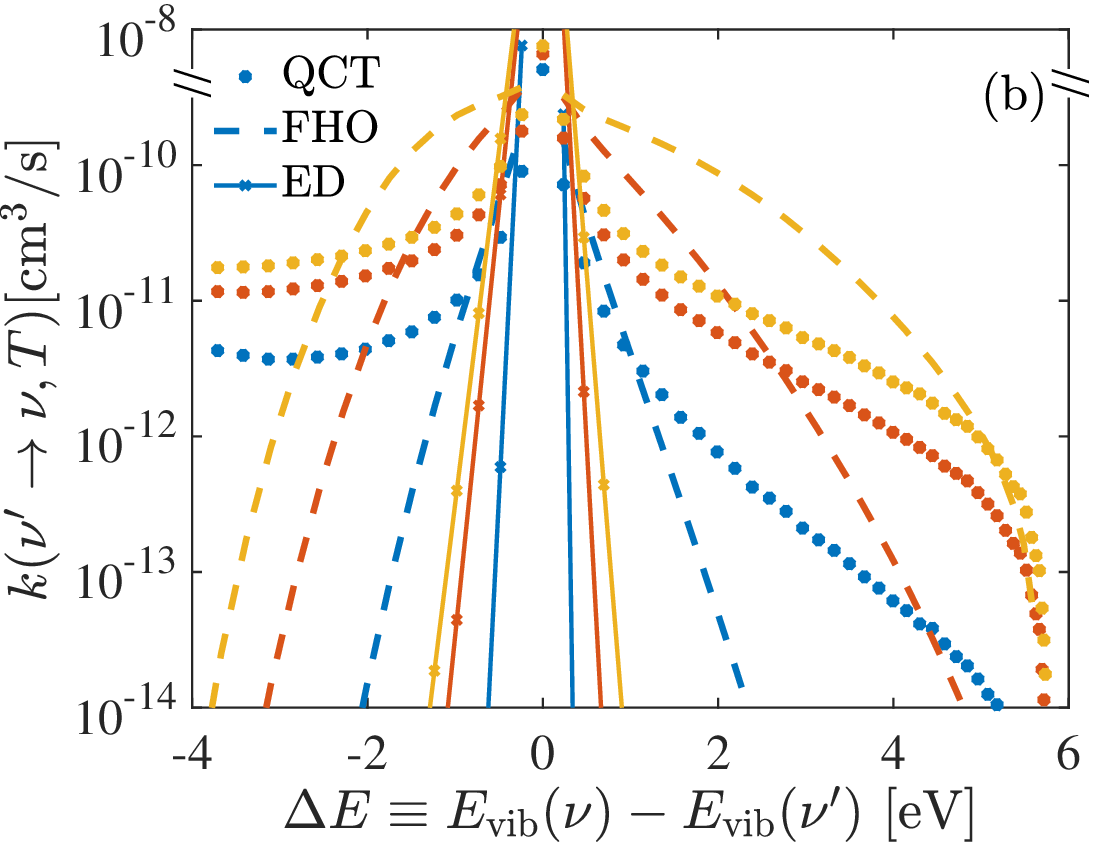}
  \end{subfigure}
\caption{
  Transition rate coefficients $k(\nu' \rightarrow \nu , T)$ from QCT data.  
  (a) $T=$ 20000 K and $\nu'=$ 5, 15, 25, and 35. 
  (b) $T= 10000, 20000$, and 30000 K and $\nu' = 15$. The prediction results from the FHO (dashed lines) model and the exponential down (ED) model (solid lines) are also given for comparison. Blue, orange and yellow colors represent $T= 10000, 20000$, and 30000 K, respectively. }
\end{figure}

Figure 1(a) shows the transition rate coefficients $k(\nu' \rightarrow \nu , T)$ obtained using QCT starting from four different initial vibrational levels at $T=\unit[20000]{K}$, plotted as a function of the vibrational-energy change after the collision $\Delta E \equiv E_{\textrm{vib}}(\nu) - E_{\textrm{vib}}(\nu')$. 
The branches of $\Delta E > 0$ (the upward energy transfer branches), corresponding to the events of diatoms gaining vibrational energy from the collision, share common features in the four cases with different $\nu'$: 
(i) For small $\Delta E$, i.e., the energy transfer between nearby states, $k(\nu' \rightarrow \nu , T)$ decreases rapidly;
(ii) For very large $\Delta E$, the vibrational energy after the collision, $E_{\textrm{vib}}(\nu)$ becomes comparable with the disassociation energy, and thus the disassociation of the diatom becomes significant, which effectively sets up a barrier for energy transfer, and $k$ vanishes abruptly; and
(iii) For the intermediate range of $\Delta E$, the data among different $\nu'$ display interesting asymptotic self-similarity that becomes more evident when the range of realizable $\Delta E$ is wider, which means that $k(\nu' \rightarrow \nu , T)$ for $\vert \nu - \nu' \vert > 1$ are significant. This phenomenon has been reported as ``super-collision'' \cite{Oref1990,Clarke1991,Clary1995,GILBERT1995,Flynn1996,Fu2021}, and should be treated with special care.

Data at other temperatures are similar, as can be seen from Figure 1(b), which shows $k(\nu' \rightarrow \nu , T)$ vs.~$\Delta E$ at three different temperatures in the $T \sim \unit[10^4]{K}$ range, all starting from the initial vibrational state $\nu' = 15$, together with predictions from several models.
Clearly, the wide range of vibrational energy transfer cannot be explained by theoretical models that consider only energy transfer between nearby states.
A popular empirical ``exponential-down'' (ED) model that assumes that the $k$ decays exponentially with the increase of the collisional energy change \cite{Oref1990,Barker2001,Jasper2014}, although can be used to describe the the dependence of $k$ in the small $\Delta E$ range, fails to capture the trends of the data in the intermediate $\Delta E$ range. 
The more sophisticated FHO model \cite{Kerner1958,Adamovich1995_part1,Adamovich1998} achieves qualitative improvement in the sense that the predicted $k$ decreases more slowly compared to that of the exponential-down model, but still deviates considerably from the data.

Figure 1 indicates that modeling $k(\nu' \rightarrow \nu , T)$ in the intermediate range of $\Delta E$ is crucial in understanding the collisional energy transfer processes. Moreover, the asymptotic self-similar behavior of $k$ in the small and the intermediate ranges of $\Delta E$ suggests that $k(\nu' \rightarrow \nu , T)$ depends on $\Delta E$ only, rather than $\nu$ or $\nu'$. Thus we denote the transition rate coefficient as $k(\Delta E, T)$, and explore its asymptotic self-similarity in the ``upward'' ($\Delta E > 0$) process, from which the transition rate $k$ in the ``downward'' ($\Delta E < 0$) process can be obtained from the principle of detailed balance \cite{Colonna2008,Kim2013,Panesi2013,Andrienko2016}. 
We note that $k(\Delta E, T)$ can be written as
\begin{equation}\label{eq:k}
  k(\Delta E , T) = Z(T) \int P(\Delta E, E_{\rm col}) \, f_B(E_{\rm col},T) \, {\rm d} E_{\rm col}
\end{equation}
where $Z(T)$ is the well-known collision frequency, 
$f_{B}(E_{\rm col}, T)$ is the normalized Boltzmann distribution of the collisional energy $E_{\rm col}$ at temperature $T$, 
and $P(\Delta E, E_{\rm col})$ is the probability for the diatom to gain an amount of $\Delta E$ in vibrational energy from a collision with an atom with collisional energy $E_{\rm col}$.
From Eq.~\eqref{eq:k}, it is clear that the properties of $k(\Delta E, T)$ are primarily determined by the transition probability.

%
%
\begin{figure}\label{fig:P} 
  \centering
  \includegraphics[width=0.6\columnwidth]{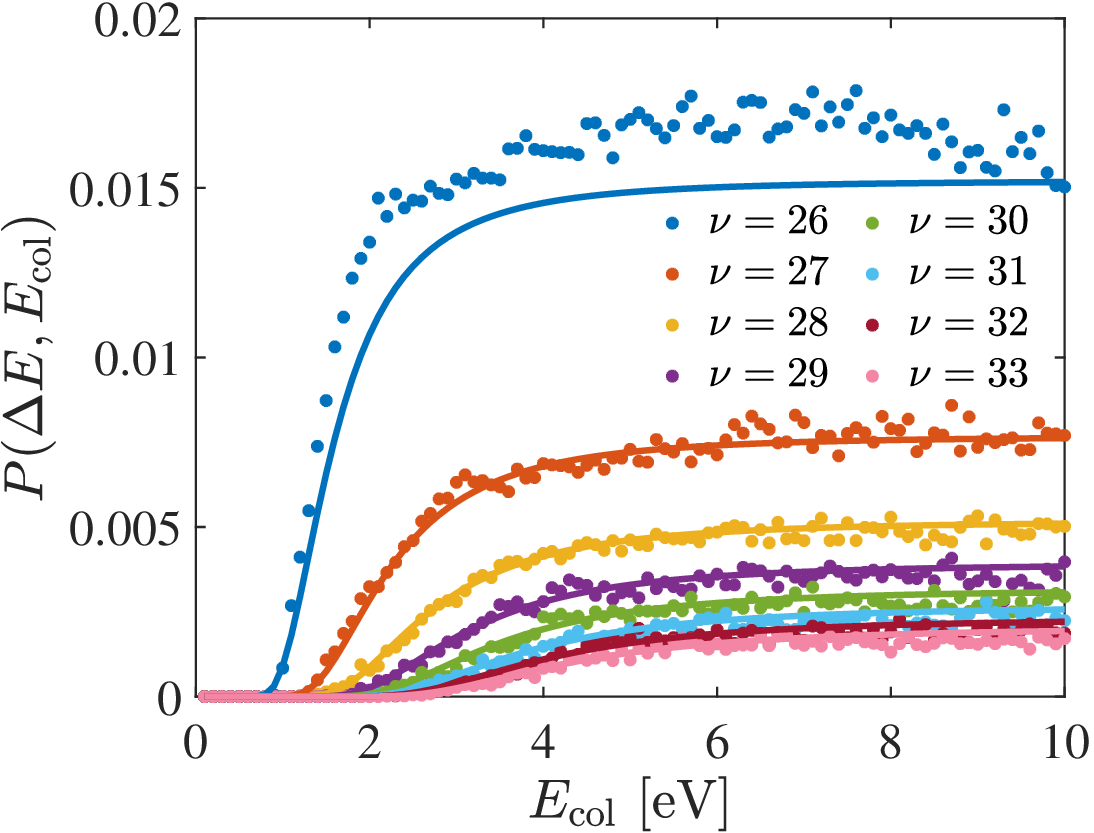}
  \caption{
Transition probability $P(\Delta E, E_{\rm col})$ 
from the initial state at $\nu' = 25$ to a specific higher-energy vibrational state $\nu$ as a function of collisional energy $E_{\rm col}$. Symbols are QCT data. Dashed lines are the fits given by Eq.~\eqref{eq:Pfit}.
}
\end{figure}
%
%

Figure \ref{fig:P} shows QCT data of the ``upward'' transition probabilities $P(\Delta E, E_{\rm col})$ from the initial state at $\nu' =25$
as a function of the collision energy $E_{\rm col}$. 
For any given state difference $\nu - \nu'$, or energy jump $\Delta E$, $P(\Delta E, E_{\rm col})$ exhibits qualitatively similar behavior:
When the collision energy $E_{\rm col}$ is low, the transition probability remains zero, which means that hardly can any increase in vibrational energy occur during collisions with low energies. As $E_{\rm col}$ increases, transitions to the nearby high vibrational level first appear. This ``activation'' gradually extends to jumps to higher and higher vibrational levels. Moreover, with the increase of $E_{\rm col}$, the transition probability to any particular higher vibrational level rapidly reaches a plateau, or ``saturation''.

Inspired by the Arrhenius form of reaction rate constants, the variation of $P(\Delta E, E_{\rm col})$ with the collision energy $E_{\rm col}$ could be described by the following empirical model:
\begin{equation}
  P(\Delta E, E_{\rm col}) = A \, \frac{E_0}{\Delta E} \, e^{ -\left( \frac{\Delta E}{E_{\rm col}} \right)^n \left( \frac{E_0}{E_{\rm col}} \right)^m}
\label{eq:Pfit}
\end{equation}
where $A$ is a dimensionless constant, $E_0$ is an energy parameter of the system,  $\Delta E$ behaves as an activation energy, and the exponents $n$ and $m$ determine how fast the activation changes to saturation -- the larger they are, the steeper the slope of the modelled $P(\Delta E, E_{\rm col})$, with very large positive values corresponding to a Heaviside-like sharp jump.
For our system, the model Eq.~\eqref{eq:Pfit} with $n\approx m\approx 1.5$, $E_0\approx 10.0 $ eV, and $A\approx 3 \times 10^{-4}$ agrees very well with the QCT data, as shown in Fig.~2.

The transition rate coefficient $k(\Delta E , T)$ can be evaluated by submitting Eq.~\eqref{eq:Pfit} in to Eq.~\eqref{eq:k}. 
Note that $P(\Delta E, E_{\rm col})$ rises up to a plateau with $E_{\rm col}$, 
while $F_{B}(E_{\rm col}, T)$ decays exponentially with $E_{\rm col}$ at any given temperature $T$, thus their product reaches the maximum 
at some intermediate $E_{\rm col}$ in the ``activation'' range and the integral in Eq.~\eqref{eq:k} is dominated by the contribution around that maximum since the product of $P(\Delta E, E_{\rm col})$ and $F_{B}(E_{\rm col}, T)$ drops rapidly on both sides. 
We thus follow Landau \cite{Landau1936} to evaluate the integral using the steepest descent integration as
\begin{equation}
  k(\Delta E , T) = Z(T) \alpha \eta^{\frac{3}{2}x-1} \zeta^{\frac{3}{2}y+1} \exp \left[ - \beta \eta^{x} \zeta^{y} \right]
\label{eq:kapprox}
\end{equation}
where $\eta \equiv \frac{\Delta E}{k_{\rm B}T}$, $\zeta \equiv \frac{E_0}{k_{\rm B}T}$, the fraction power $x=\frac{n}{n+m+1}$ and $y=\frac{m}{n+m+1}$, $k_{\rm B}$ is the Boltzmann constant, $\alpha$ and $\beta$ are numerical factors, see SM for details. Later we refer Eq.~\eqref{eq:kapprox} as the activation-saturation (AS) model.

\begin{figure}[htbp]
  \centering
  \includegraphics[width=0.6\columnwidth]{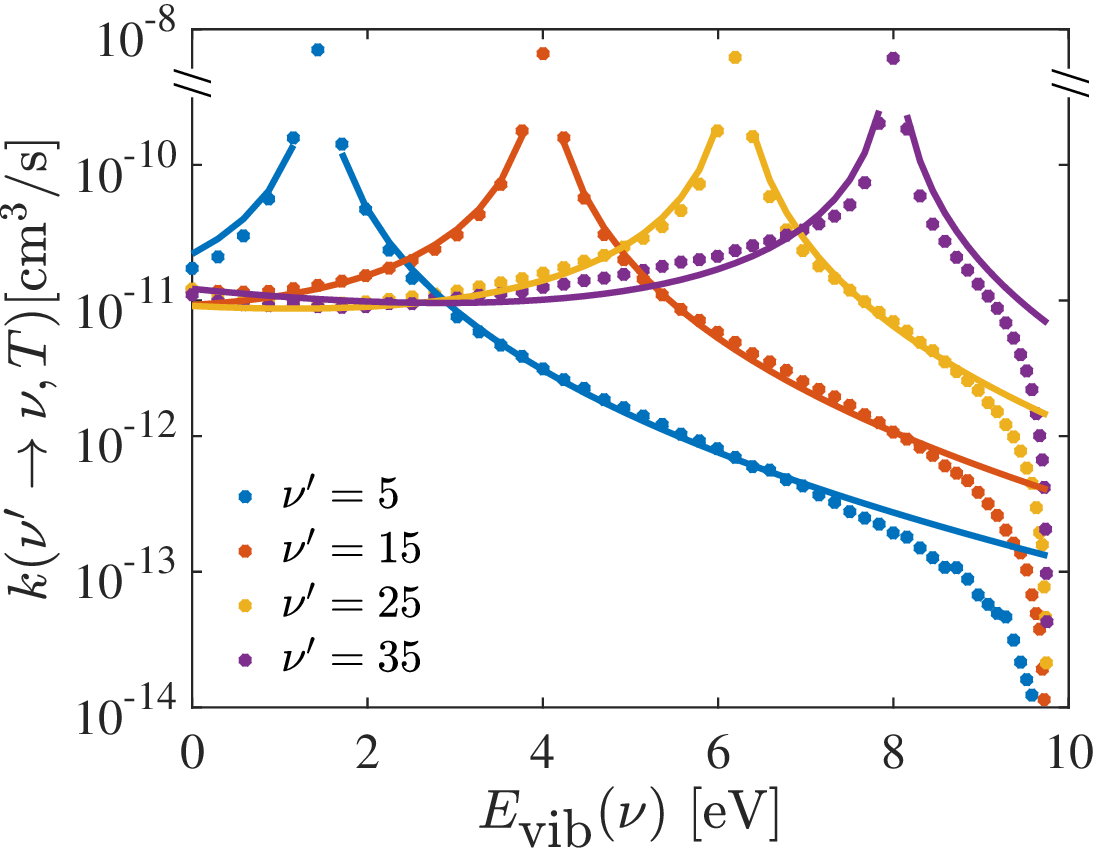}
  \caption{
  Transition rate coefficients $k(\nu' \rightarrow \nu , T=\unit[20000]{K})$ vs. $E_{\rm vib}(\nu)$, the vibrational energy of the final state. 
  Symbols are QCT data and solid-lines are Eq.~\eqref{eq:kfit}. 
  }
  \label{fig:kEvib}
\end{figure}

Interestingly, Eq.~\eqref{eq:kapprox} shows that the widely used exponential-down form $k(\Delta E , T) \sim \exp(- \Delta E)$ corresponds to $n \gg m$ and $n \gg 1$, which, according to Eq.~\eqref{eq:Pfit}, suggests that the transition probability reduces to a Heaviside function. This reveals an implicit assumption behind the exponential-down form for the collisional energy transfer.
More generally, $k(\Delta E , T)$ depends on the energy gap in the form of stretched exponential, $(\Delta E)^\frac{n}{n+m+1}$.
For the N+N$_2$ system tested, $n \approx m \approx 1.5$, the transition rate coefficient in the ``upward energy transfer''  given by the AS model becomes 
\begin{equation}
  k(\Delta E , T)  =  Z(T) \alpha \eta^{-\frac{7}{16}} \zeta^{\frac{25}{16}} \exp \left[ - \beta \eta^{\frac{3}{8}} \zeta^{\frac{3}{8}} \right]
\label{eq:kfit}
\end{equation}
in which the numerical factors $\alpha=0.0036$ and $\beta=2.12$ are determined from fitting the QCT data.
As shown in Fig.~\ref{fig:kEvib}, the results agree well with the QCT data over a wide range of energy transfer.

The AS model Eq.~\eqref{eq:kapprox} succinctly and accurately describes the energy transfer across wide gaps of energy levels. 
This ``long-range'' energy transfer 
plays an important role in the evolution of the vibrational energy distribution at high temperatures, 
which has been reported before as the ``super-collision'' effect and a ``double-exponential'' model \cite{Oref1990} was proposed 
to improve the traditional exponential-down model. 
Here, the AS model shows that the transition rate coefficients follow a single stretched-exponential form in the entire range of vibrational energies, which also explains the observation from earlier experiments \cite{Hold2000,Lenzer2000}.
This behavior is not only valid for the N+N$_2$ system, the same has been observed 
in our simulation with 
O+N$_2$, and O+O$_2$ systems, with only small changes in the numerical values of the parameters (see SM).

\begin{figure}[htbp]
  \begin{subfigure}
    \centering
    \includegraphics[width=0.48\columnwidth]{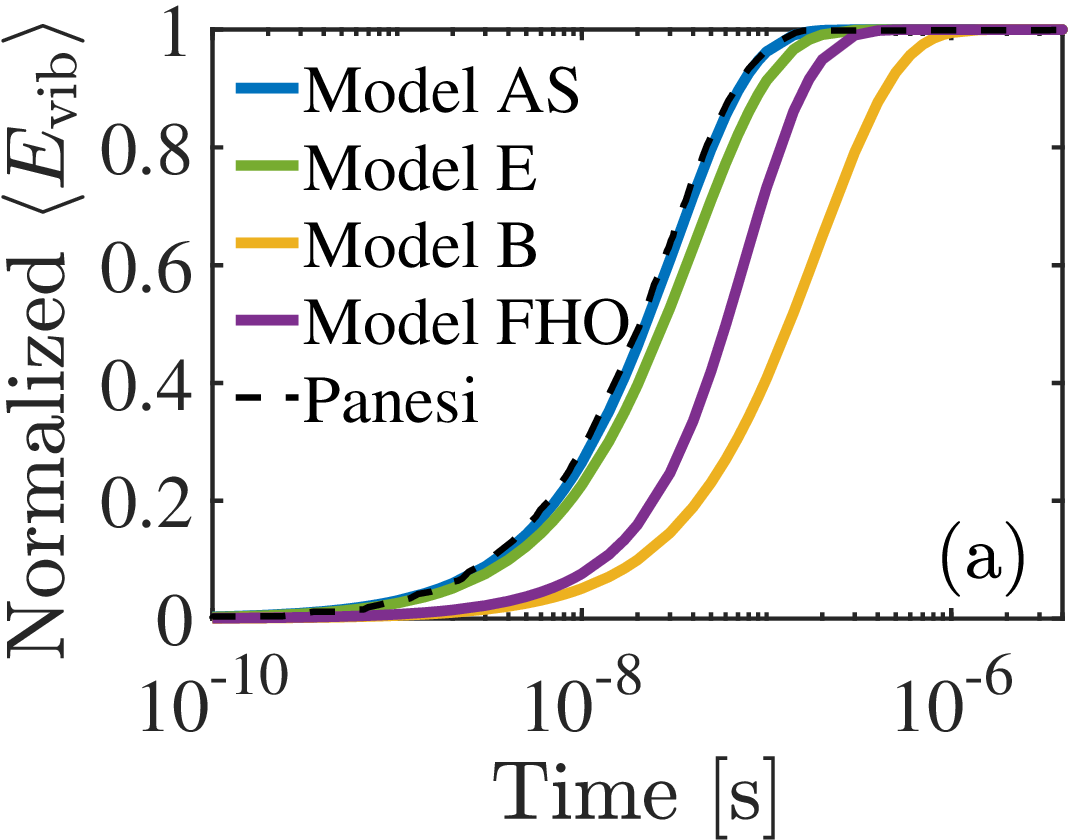}
  \end{subfigure}
  \hfill
  \begin{subfigure}
    \centering
    \includegraphics[width=0.48\columnwidth]{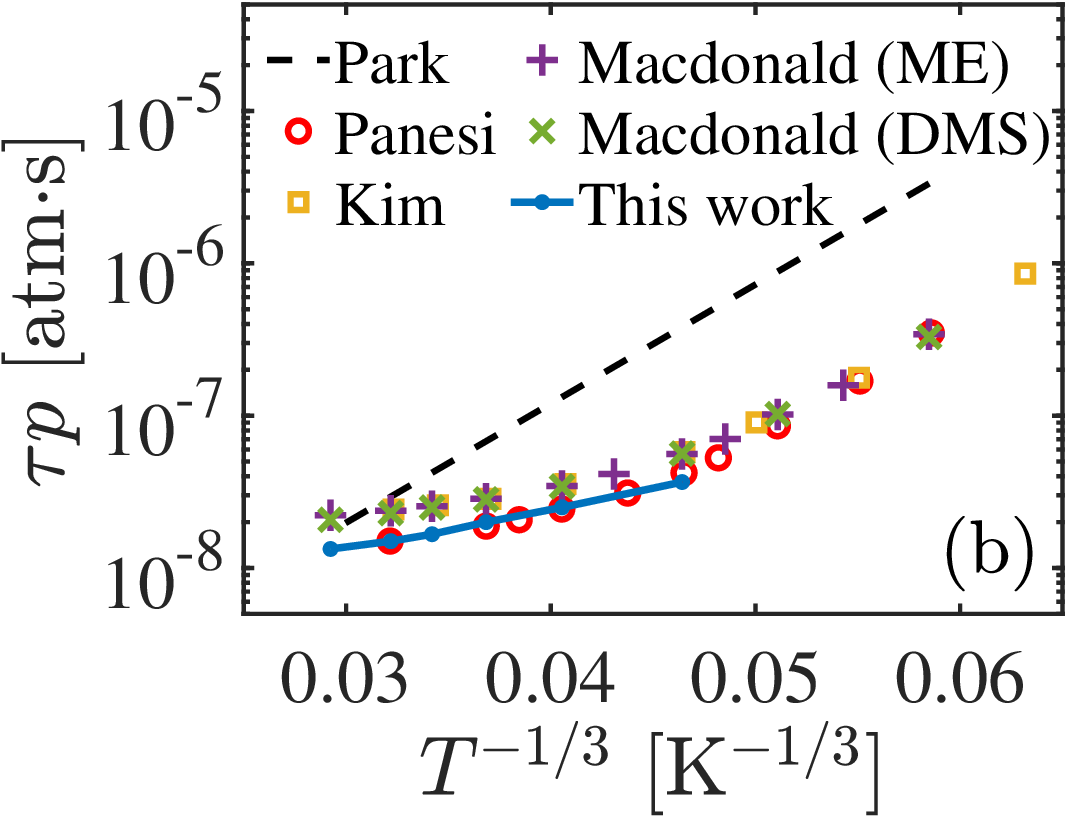}
  \end{subfigure}
\caption{\label{fig:Etime}
Relaxation of the vibrational energy of N$_2$ in the N+N$_2$ mixture after passing a strong shock. 
(a) Change of the normalized vibrational energy 
with time at translational temperature $T=\unit[30000]{K}$. 
(b) The e-folding time $\tau$ vs. $T$.
Symbols are data from the literature: $\circ$ Panesi et al. \cite{Panesi2013}, $\square$ Kim et al.~\cite{Kim2013}, $+$ and $\times$ are the ME and the direct molecular simulation (DMS) results from MacDonald et al.~\cite{Macdonald2020}, dashed-line is Park's model \cite{Park1993}, and solid-line is from this work.
}
\end{figure}

With the transition rate coefficients given by the AS model, the evolution of the vibrational energy distribution in a gas undergoing non-equilibrium processes can be obtained by solving the master equations (ME). As an example, we consider a mixture of N+N$_2$ in which the initial vibrational temperature of N$_2$ is at a temperature $T_{\rm v}$ much lower than the translational temperature $T$ of both N and N$_2$, which is a model system to mimic the temperature non-equilibrium of N$_2$ gas after passing a strong shock. This model has been simulated using the master equation approach \cite{Panesi2013} with the state-to-state transition rate coefficients carefully tabulated in the NASA database \cite{Jaffe2010}. In the simulation \cite{Panesi2013}, the gas mixture was kept at a constant number density of $\unit[2.4\times10^{18}]{cm^3}$ with the molar fraction of the N atom in the mixture fixed at 5\%.
The initial vibrational temperature was set to be $T_{\rm v} = \unit[300]{K}$, which corresponds to the temperature of the gas before encountering the shock. 
The strength of the shock, or the Mach number, was reflected by the translational temperature $T$ of the gas and was fixed during the simulation.
Figure 4(a) shows the change of the normalized total vibrational energy $\left \langle E_{\rm vib} (t) \right \rangle / \left \langle E_{\rm vib} (t \to \infty) \right \rangle$ with time $t$ in the case of $T = \unit[30000]{K}$. It is clear that for the high-translational-temperature case simulated, the AS model result is the closest to the benchmark given in Ref.~\cite{Panesi2013}, comparable or slightly better than those from the empirical fits in Ref.~\cite{Esposito2006} (Model E) that contains several hundreds of fitting parameters, while the model by Billing \& Fisher \cite{Billing1976} (Model B) and the FHO model predict much slower relaxation of the vibrational energy.
Similar agreements are observed for cases at other temperatures in the range of $T \sim \unit[10^4]{K}$ (see SM). 

The relaxation process is commonly characterized by the e-folding time $\tau$ of the vibrational energy \cite{Park2004}. 
An influential model by Park \cite{Park1993} based on the empirical equation proposed by Millikan \& White \cite{Millikan1963} has the form of $\ln(\tau p) \sim a T^{-1/3} -b$, where $p$ is the pressure of the bath gas (N atoms in the current work), 
$a$ and $b$ are parameters fitted from experimental data. 
Figure 4(b) shows that the dependence of $\ln(\tau p)$ on $T^{-1/3}$ obtained with the transition rate coefficients 
given by the AS model is satisfactory in the high temperature range, within a factor of two of all the available simulation data \cite{Panesi2013,Kim2013,Macdonald2020}, where the discrepancy can be attributed to our simplified treatment of the rotational energy.
While Park's model captures the trends in the high and low temperature range separately, 
and the slope $a$ given in the original Park's model \cite{Park1993} is consistent only with the data in the low-temperature range (see SM for more discussion). 
The AS model thus could help to investigate the relaxation time in the high temperature range.

\section*{Acknowledgments}
We are grateful to the Natural Science Foundation of China for financial support through grants 11988102 and 21973053.

\renewcommand{\bibfont}{\footnotesize\linespread{0.9}\selectfont}
\bibliography{references}
\newpage



\end{document}